\documentclass[twocolumn,amstext,amsmath,amssymb,prc,superscriptaddress,floatfix,showpacs,nofootinbib]{revtex4}

\usepackage[dvips]{epsfig}
\usepackage{slashed}
\usepackage{mathrsfs}
\usepackage{graphicx}
\usepackage{xcolor}
\usepackage{hyperref}
\usepackage{dcolumn}
\usepackage{subfigure}
\usepackage{xspace}

\newcommand{\mm}{\marginpar{\colorbox{green}{\textbf{BJ}}\\@Mario:}}
\newcommand{\bj}{\marginpar{\colorbox{green}{\textbf{MM}}\\@BJ:}}

\def\roughly#1{\mathrel{\raise.3ex\hbox{$#1$\kern-.75em%
\lower1ex\hbox{$\sim$}}}}

\def\g2k{\Gamma^{(2)}_k}

\def\ma0{m_{a_{0}}}
\def\mf0{m_{f_{0}}}

\definecolor{heidelbeer}{rgb}{0.5,0,0.5}

\graphicspath{{./Figures/}
}
\begin{document}
\title{Optimizing the pulse shape for Schwinger pair production}

\author{C. Kohlf\"{u}rst}
 \email[]{christian.kohlfuerst@uni-graz.at}
 \affiliation{Institut f\"{u}r Physik, Karl-Franzens-Universit\"{a}t,
   A-8010 Graz, Austria}

 \author{M. Mitter}
 \email[]{mario.mitter@uni-graz.at}
  \affiliation{Institut f\"{u}r Physik, Karl-Franzens-Universit\"{a}t,
   A-8010 Graz, Austria}

\author{G. von Winckel}
 \email[]{gregvw@gmail.com}
 \affiliation{Institut f\"{u}r Mathematik und wissenschaftliches Rechnen, Karl-Franzens-Universit\"{a}t, A-8010 Graz, Austria}
 \affiliation{Department of Electrical and Computer Engineering, University of New Mexico, Albuquerque-NM 87106, USA}

 \author{F. Hebenstreit}
 \email[]{f.hebenstreit@thphys.uni-heidelberg.de}
  \affiliation{Institut f\"{u}r Theoretische Physik, Universit\"{a}t Heidelberg,
   D-69120 Heidelberg, Germany}

  \author{R. Alkofer}
 \email[]{reinhard.alkofer@uni-graz.at}
  \affiliation{Institut f\"{u}r Physik, Karl-Franzens-Universit\"{a}t,
   A-8010 Graz, Austria}

\date{\today}

\begin{abstract}
Recent studies of the dynamically assisted Schwinger effect have shown that particle production is significantly enhanced by a proper choice of the electric field. 
We demonstrate that optimal control theory provides a systematic means of modifying the pulse shape in order to maximize the particle yield.
We employ the quantum kinetic framework and derive the relevant optimal control equations.
By means of simple examples we discuss several important issues of the optimization procedure such as constraints, initial conditions or scaling.
By relating our findings to established results we demonstrate that the particle yield is systematically maximized by this procedure.
\end{abstract}

\pacs{12.20.Ds, 11.15.Tk, 02.60.Pn}
\maketitle

\section{Introduction}
\label{sec:intro}

Creation of electron-positron pairs by a strong and (quasi-)static electric field, the so-called Schwinger effect, has been a long-standing but still unobserved prediction of early quantum theory \cite{Sauter:1931zz,Heisenberg:1935qt,Schwinger:1951nm}. 
All attempts to observe this fundamental, non-perturbative effect of strong-field QED have failed due to its exponential suppression up to an unprecedented field strength of $E_c\sim10^{18}\,\mathrm{V/m}$.
The rapid development of laser technology in recent years, however, has raised the hope to produce peak field strengths of this order such that a direct observation of the Schwinger effect might become possible soon \cite{Ringwald:2001ib}. 

Another possibility to create electron-positron pairs is to supply the necessary rest mass energy by an arbitrarily weak external field with frequency $\omega>2m$, with $\hbar=c=1$ throughout.
Remarkably, this production mechanism works even for frequencies $\omega<2m$ by an absorption of multiple photons, however, it becomes strongly suppressed in this regime \cite{Brezin:1970xf,Popov:1972,Alkofer:2001ik}.
As a matter of fact, the multiphoton pair production mechanism has been used to explain the outcome of the SLAC E--144 experiment more than a decade ago \cite{Bamber:1999zt}.

A few years ago, it was suggested to employ multiphoton pair production to assist the Schwinger effect, referred to as dynamically assisted Schwinger mechanism \cite{Schutzhold:2008pz}.
Note that similar proposals on how to lower the threshold for pair production with the aid of more advanced setups have been put forward as well \cite{DiPiazza:2009py,Dunne:2009gi,Monin:2009aj,Bulanov:2010ei}. 
The idea of the dynamically assisted Schwinger \mbox{mechanism} has been pursued recently and it has been shown that a proper parameter choice can in fact result in an enhancement of the particle yield by orders of magnitude \cite{Orthaber:2011cm,Fey:2011if}. 
However, all these studies have been based on rather simple field configurations and first attempts to investigate more sophisticated ones have been made only recently \cite{Hebenstreit:2009km,Dumlu:2010vv,Akkermans:2011yn,Kohlfurst:2012,Nuriman2012465}.
The parameter space (field strengths, time scales, etc.), however, grows rapidly for more complicated field configurations so that a systematic treatment becomes impracticable very soon.

In order to overcome this drawback, we will apply optimal control theory to the problem of electron-positron pair creation for the first time. 
Note that this optimization method has already been widely used in several other areas of physics such as molecular physics \cite{PhysRevA.37.4950}, quantum computing \cite{PhysRevLett.89.188301}, or splitting of Bose-Einstein condensates \cite{PhysRevA.80.053625}, to name only a few.
We will employ this method in order to systematically shape the electric field under certain constraints such that the number of created electron-positron pairs gets maximized.
In this respect, a proper choice of boundary conditions will be crucial.

Recent experimental prospects have indicated that the available field strengths at the European XFEL as well as the Extreme Light Infrastructure (ELI) will still be sub-critical.
Accordingly, a direct observation of the Schwinger effect will most probably depend on an optimal choice of parameters. 
Such optimized field configurations have already been used in atomic ionization, which is rather similar to the Schwinger effect from a theoretical point of view, 
showing that the ionization rate can be significantly enhanced by an optimized pulse shape \cite{PhysRevLett.92.208301}.  
In this respect it is worth mentioning that the shaping of femtosecond pulses is an available cutting-edge technique which allows for the generation of complicated field configurations according to user specification \cite{weiner:1929}.

The structure of this paper is as follows: 
In section \ref{sec:motivation} we briefly introduce the quantum kinetic framework and present recent results on the dynamically assisted Schwinger mechanism.
This section will motivate the necessity of a systematic pulse-shaping procedure in order to maximize the particle yield.
In section \ref{sec:theory} we introduce optimal control theory and apply it to the quantum kinetic formalism. 
We present the corresponding numerical results in section \ref{sec:results}. 
Finally, we conclude and give an outlook in section \ref{sec:conclusion}.

\section{Motivation}
\label{sec:motivation}

Our electromagnetic field choice is motivated by the focal region of counter propagating laser pulses.
Given that the spatial variation scale is much larger than the \mbox{electron's} Compton wavelength all spatial inhomogeneities may be ignored \cite{Hebenstreit:2011wk}.
Moreover, assuming a standing wave mode we may neglect all magnetic effects as well, $\mathbf{B}(t)=0$. 
Accordingly, we will only investigate electron-positron pair creation in the presence of a spatially homogeneous, time-dependent electric field $\mathbf{E}(t) = E(t)\,\mathbf{e}_3$. 
We will represent this field configuration in terms of a time-dependent vector potential $\mathbf{A}(t)=A(t)\,\mathbf{e}_3$ in temporal gauge $A_0=0$, so that:
\begin{equation}
 \mathbf{E}(t)=-\dot{\mathbf{A}}(t) \qquad \mathrm{and} \qquad \mathbf{B}(t) = \nabla\times\mathbf{A}(t)= 0 \ .
\end{equation}

\subsection{Quantum kinetic formalism}
\label{sec:motivation_qk}

In order to calculate electron-positron pair creation in this type of electric background field, we will employ the quantum kinetic formalism. 
Within this approach, all spectral information is encoded in the distribution function $F(\mathbf{q},t)$.
We denote the canonical momentum parallel to the field direction as $q$ and restrict ourselves to $\mathbf{q}_\perp=0$.
Correspondingly, we will have for the transverse energy squared $\epsilon_\perp^2=m^2+\mathbf{q}_\perp^2\rightarrow m^2$.
This simplification is justified as particle production occurs dominantly at small orthogonal momenta, even if a distinctive cellular structure in momentum space can be observed for complicated field configurations \cite{Blaschke:2012vf}. 

As particle production is a genuine non-equilibrium problem in quantum field theory, it is clear that a particle interpretation of the distribution function is only valid at asymptotic times $T$\footnote{Here a remark with respect to the recent investigation presented in Ref.~\cite{Nuriman2012465} is in order.
We want to point out that presenting momentum spectra at peak field values might be misleading as the distribution function can be related to the particle yield only at asymptotic times and vanishing fields (as has been acknowledged by the authors themselves).
In addition, based on our own simulations, we suspect that some of their results suffer from problems due to a lack of accuracy in the numerical computation.}.
Accordingly, the momentum space integral  
\begin{equation}\label{eq:density}
  n[F]=\int\limits_{-\infty}^\infty{F(q,T)dq} 
\end{equation}
can be interpreted as the density of created electron-positron pairs for $\mathbf{q}_\perp=0$.

The time evolution of the system is governed by an integro-differential equation \cite{Schmidt:1998vi} or, from a numerical perspective more conveniently, an ordinary differential equation system \cite{Bloch:1999eu}:
\begin{subequations}
 \label{eq:qke}
 \begin{align}
 \dot{F}(q,t) & =  W(q,t)G(q,t) \ ,  \\
 \dot{G}(q,t) & =  W(q,t)[1-F(q,t)]-2\omega(q,t) H(q,t) \ ,  \\
 \dot{H}(q,t) & =  2\omega(q,t) G(q,t) \ , 
 \end{align}
\end{subequations}
where the dot denotes the derivative with respect to $t$.
Here, $G(q,t)$ and $H(q,t)$ are auxiliary functions and the remaining quantities are defined according to:
\begin{equation}
 \omega^2(\mathbf{q},t) = \epsilon_\perp^2 + [q - e A(t)]^2 \ \ , \ \ W(q,t) = \frac{eE(t)\epsilon_\perp}{\omega^2(q,t)} \ .
\end{equation}
The initial conditions in the asymptotic past are given by $F(q,-T)=G(q,-T)=H(q,-T)=0$.
Note that we neglect the back reaction effect of created pairs on the electric field which is expected to be a good approximation in the sub-critical field strength regime.

\begin{figure}[t]
  \centering
  \includegraphics[width=\columnwidth]{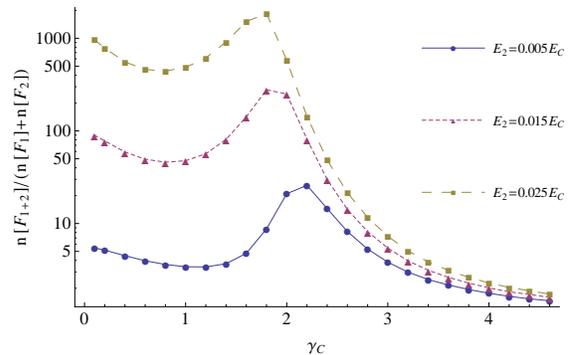}
  \caption{\label{fig:enhancement} 
  Enhancement of the particle density in the dynamically assisted Schwinger mechanism. 
  The parameters of the adiabatic pulse are given by $E_1=0.1E_c$ and $\omega_1\sim m/100$. 
  The different curves correspond to different values of $E_2$ and we change $\gamma_c$ or, equivalently, $\omega_2$.}
\end{figure}

\subsection{Dynamically assisted Schwinger mechanism}

Based on the quantum kinetic formalism, we are able to calculate the momentum spectrum as well as the number of created electron-positron pairs for arbitrary field configurations.
As mentioned previously, one type of field configuration which has attracted much interest recently consists of a superposition of a strong adiabatic pulse (index 1) with a weak anti-adiabatic pulse (index 2):
\begin{equation}
 \label{eq:two_pulses}
 E(t)=E_1\operatorname{sech}^2(\omega_1 t)+E_2\operatorname{sech}^2(\omega_2 t) \ .
\end{equation}
The adiabaticity of a single pulse is quantified in terms of the Keldysh parameter:
\begin{equation}
 \gamma=\frac{m\omega}{eE} \ .
\end{equation}
This parameter discriminates between two different regimes: The pair creation process shows the characteristics of the Schwinger effect for $\gamma \ll 1$ whereas the multiphoton regime is associated with $\gamma \gg 1$.
For the field configuration (\ref{eq:two_pulses}) we assume $\gamma_1 \ll 1$ and $\gamma_2 \gg 1$ or, equivalently, $E_c \gg E_1 \gg E_2$ and $\omega_1 \ll \omega_2 \ll m$.
It has been pointed out, however, that neither $\gamma_1$ nor $\gamma_2$ is the relevant parameter when dealing with the dynamically assisted Schwinger effect but rather a combined Keldysh parameter, which is composed of the dominant scales \cite{Schutzhold:2008pz}:
\begin{equation}
 \gamma_c=\frac{m\omega_2}{eE_1} \ .
\end{equation}
In fact, the non-trivial interplay between different scales results in a subtle change of the pair production characteristics for properly chosen parameters.
Most notably, it has been observed that $F_{1+2}(q,T)$ shows strong deviations compared to $F_1(q,T)+F_2(q,T)$ \cite{Orthaber:2011cm}.
Consequently, this results in a strong enhancement of the particle yield, i.~e. $n[F_{1+2}]$ can become orders of magnitude larger than $n[F_1]+n[F_2]$.

To illustrate this, we superimpose an adiabatic pulse with parameters $E_1=0.1E_c$ and $\omega_1 \sim m/100$ with an anti-adiabatic pulse for different values of $E_2$ as function of $\gamma_c$ in FIG.~\ref{fig:enhancement}.
It can be seen that the enhancement peaks around $\gamma_c=0$ as well as $\gamma_c\sim2$.
The case $\gamma_c=0$ corresponds to a trivial effect as the second pulse turns from the anti-adiabatic to the adiabatic regime.
Consequently, particle creation becomes enhanced simply due to a higher peak field strength $E_1+E_2$.
The peak around $\gamma_c\sim2$, however, is a nontrivial result of the analysis of the dynamically assisted Schwinger mechanism.
Finally, the enhancement approaches unity for large values of $\gamma_c$.
In this case, the second pulse completely dominates as it approaches the perturbative pair creation threshold.

\begin{figure}[t]
  \centering
  \includegraphics[width=0.92\columnwidth]{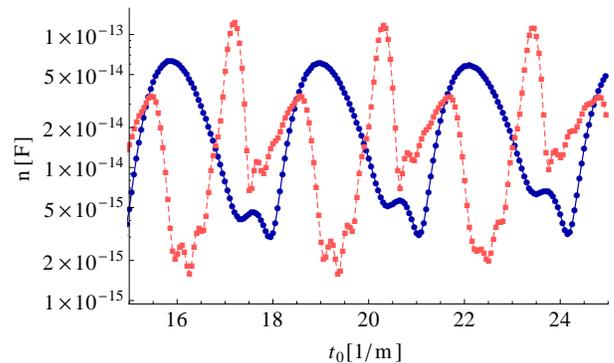}
  \caption{\label{fig:comb} 
  Particle density for of a comb of $10$ single pulses with parameter $|E_i|=0.02E_c$, $\omega_i=m/6$ for the equal-sign configuration (solid) and the alternating-sign configuration (dashed). 
  The particle number changes quasi-periodically by orders of magnitude as function of the inter-pulse time lag $t_0$.}
\end{figure}

This simple example nicely illustrates the general idea of the dynamically assisted Schwinger mechanism.
However, it is already rather challenging to investigate the whole four-dimensional parameter space $\{E_1,E_2,\omega_1,\omega_2\}$.
This becomes even worse when we consider more complicated field configurations.
For instance, we may generalize (\ref{eq:two_pulses}) to a superposition of an arbitrary number of single pulses as investigated recently \cite{Akkermans:2011yn}:
\begin{equation}
 \label{eq:spulse_sum}
 E(t)=\sum_{i}{E_i\operatorname{sech}^2(\omega_i[t+t_{0,i}])} \ .
\end{equation}
In this case, all pulses may have a different field strength $E_i$, frequency $\omega_i$ and time lag $t_{0,i}$. 
Accordingly, a systematic investigation becomes absolutely impracticable.

In order to get an impression of the complicated structure of the parameter space, we employ the idea presented in Ref.~\cite{Akkermans:2011yn} where it has been shown that a sequence of alternating-sign pulses produces a Ramsey interferometer.
Based on this observation, we investigate the alternating-sign configuration as well as the equal-sign configuration in FIG.~\ref{fig:comb}.
To be specific, we show the number of created particles for a comb of $10$ single pulses with parameters $|E_i|=0.02E_c$, $\omega_i=m/6$ and $t_{0,i}=i \cdot t_0$ as function of the inter-pulse time lag $t_0$.
Note that we have $\operatorname{sign}[E_i]=+1$ for the equal-sign configuration whereas we have $\operatorname{sign}[E_i]=\pm1$ in the alternating-sign configuration for $i$ being even and odd, respectively.

We observe some remarkable features upon varying the inter-pulse time lag $t_0$. 
On the one hand, the number of created particles changes quasi-periodically by orders of magnitude for both the alternating-sign and the equal-sign configuration.
This meets the expectations for the alternating-sign configuration according to the discussion in Ref.~\cite{Akkermans:2011yn}, however, the appearance of a pronounced interference effect for the equal-sign configuration has not been anticipated.
On the other hand, the local extrema in the particle density show an out-of-phase behavior for these two configurations. 
A more detailed investigation and discussion of the differences between the alternating-sign and equal-sign configurations can be found in Ref.~\cite{Kohlfurst:2012}.
The quasi-periodic structure of the pronounced maxima strongly suggests some kind of resonance effect which is triggered by the inter-pulse time lag $t_0$. 
This clearly demonstrates the sensitivity of the particle yield on rather minor parameter changes besides the field strength. 
A more profound explanation of this effect is beyond the scope of this manuscript, however, 
effects due to multiphoton absorption could probably account for the observed interferences \cite{MultiPhoton}.

\section{Pulse shaping}
\label{sec:theory}

We have seen in the previous section that a proper parameter choice is crucial for maximizing the particle yield.
However, we have also observed that a systematic scan of the parameter space becomes rather impracticable for field configurations more complicated than (\ref{eq:two_pulses}).
In order to overcome this drawback, we will now introduce optimal control theory to increase the particle yield in a more systematic way.

\subsection{Optimal Control}
\label{sec:theory_optcontrol}

The task of maximizing the number of produced electron-positron pairs by varying the field configuration can be formulated as an optimization problem with constraints.
Moreover, it is obvious from a physical point of view that it is not sufficient to maximize $n[F]$ without any restrictions.
Due to the fact that the Schwinger effect is exponentially suppressed, for instance, an increase of the particle yield could be trivially achieved by increasing the field strength.
As a consequence, we have to put additional limitations on the admissible field configurations.
Accordingly, the objective functional, which is to be minimized, is taken to be:
\begin{eqnarray}
 \label{eq:cost_functional}
 J[F,A] & = & -\gamma \, n[F] + f[A]\ .
\end{eqnarray}
Here $\gamma$ is some constant to be specified, $A$ is the vector potential and the functional $f[A]$ has been added to restrict the set of admissible vector potentials $A$. 
The objective functional is subject to the constraints $e_{F,G,H}(q,t)$ corresponding to the equations of motion (\ref{eq:qke}):
\begin{subequations}
 \label{eq:eom}
 \begin{align}
  0 = e_F & =  \dot{F} - WG\ ,  \\
  0 = e_G & =  \dot{G} + WF+2\omega H - W\ ,  \\
  0 = e_H & =  \dot{H}-2\omega G\ . 
 \end{align}
\end{subequations}
Apart from the inhomogeneous contribution $W(q,t)$ to $e_G(q,t)$, these equations are determined by a skew-symmetric matrix which conserves the norm of the vector $(F,G,H)$ in time.
These constraints can be accounted for by introducing Lagrange multipliers $\mu_{F,G,H}(q,t)$ leading to the Lagrangian
\begin{alignat}{2}
 \label{eq:Lagrangian}
 L(F,&G,H,A,\mu_F,\mu_G,\mu_H)  =  \\
     &J[F,A] + \langle \mu_F,e_F\rangle_{\Omega} + \langle \mu_G,e_G\rangle_{\Omega} + \langle \mu_H,e_H\rangle_{\Omega}\ , \quad \nonumber
\end{alignat}
where $\langle \cdot,\cdot\rangle_{\Omega}$ denotes the $L^2$ inner product on $\Omega = \mathbb{R}\times[-T,T]$. 
A vanishing gradient of the Lagrangian is a necessary condition for $F,G,H,A$ to be a local minimizer of $J[F,A]$.
Setting the gradient with respect to $(\mu_F,\mu_G,\mu_H)$ equal to zero reproduces the constraint equations (\ref{eq:eom}). 

More interestingly, the gradient with respect to $(F,G,H)$ leads to the adjoint equations: 
\begin{subequations}
\label{eq:adjoint}
 \begin{align}
  & 0 = \dot{\mu}_F-W\mu_G\ ,\\
  & 0 = \dot{\mu}_G+W\mu_F+2\omega \mu_H \ ,\\
  & 0 = \dot{\mu}_H-2\omega \mu_G \ ,
 \end{align}
\end{subequations}
with final conditions $\mu_F(q,T) = \gamma$ and $\mu_G(q,T)=\mu_H(q,T)=0$. 
Apart from the inhomogeneous term these equations of motion for the Lagrange multipliers are identical to (\ref{eq:eom}).

Finally, the variation of the Lagrangian with respect to $A$ can be calculated straightforwardly as well and gives, apart from time-independent boundary terms:
\begin{widetext}
 \begin{eqnarray}
  \label{eq:gradient}
  \frac{\delta L}{\delta A} & = & \frac{\delta f}{\delta A} + e\int\limits_{-\infty}^\infty\left\{2\left(\mu_H G - \mu_G H\right)\frac{q-eA}{\omega} 
+ \frac{\epsilon_{\perp}}{\omega^2}\frac{d}{dt}\left(\mu_G F - \mu_F G - \mu_G\right)\right\}dq\ .
 \end{eqnarray}
\end{widetext}
Assuming that the equations of motion (\ref{eq:eom}) can be solved uniquely for any admissible potential $A$, it is useful to define the reduced cost functional
\begin{eqnarray} 
 \label{eq:red_cost_func}
 \hat{J}[A] & = & J[F(A),A]\ .
\end{eqnarray}
With the aid of $\hat{J}[A]$ the formerly constrained optimization problem can be reformulated as an optimization problem without constraints.
As a matter of fact, most optimization methods require at least knowledge about the gradient of the corresponding objective functional. 
In this respect, it can be shown that the gradient of the reduced cost functional with respect to the control function $A$ is given by
\begin{equation}
 \frac{\delta \hat{J}[A]}{\delta A}=\frac{\delta L}{\delta A} 
\end{equation}
if $F,G,H$ is the unique solution of (\ref{eq:eom}) and $\mu_F,\mu_G,\mu_H$ fulfill the adjoint equations (\ref{eq:adjoint}).
Accordingly, apart from solving the equations of motion forwards in time it is also necessary to evolve the adjoint equations backwards in time in order to calculate search directions for the reduced cost functional.

\subsection{Constraints}
\label{sec:constraints}

The previous derivation has been general in the sense that we have not specified the functional $f[A]$ in order to implement physical restrictions on the admissible field configurations.
As a matter of fact, the possibilities are virtually unlimited but for the sake of simplicity we will discuss only two restrictions at this point.

The first limitation concerns the electric field strength as discussed previously.
To ensure boundedness of the electric field $E=-\dot{A}$ at any time
\begin{equation}
 \label{eq:bounds}
 E_{\text{min}}(t) \leq -\dot{A}(t) \leq E_{\text{max}}(t) \ ,
\end{equation}
we implement the cost functional:
\begin{equation}
 \label{eq:bounds1}
 f_1[A] = -\gamma_1 h[-A,-E_{\text{min}}] - \gamma_2 h[A,E_{\text{max}}] \ ,
\end{equation}
with
\begin{equation}
 \label{eq:bounds2}
 h[A,E]  =  \int\limits_{-T}^{T}\log\left(E(t) + \dot{A}(t)\right)dt \ .
\end{equation}
Here $\gamma_{1,2}$ are positive numbers and both $E_{\text{min}}$ and $E_{\text{max}}$ should vanish at asymptotic times $\pm T$ to ensure an asymptotic particle interpretation in this limit.
The functional will diverge $f_1[A]\to+\infty$ if either $-\dot{A} \searrow E_{\text{min}}$ or $-\dot{A} \nearrow E_{\text{max}}$ on some subinterval of $[-T,T]$. 
Accordingly, field configurations out of the bounds (\ref{eq:bounds}) are prevented as they would render the minimization of the objective functional impossible.
Apart from time-independent boundary terms, its variation with respect to the control function is given by:
\begin{eqnarray}
 \frac{\delta f_1}{\delta A} & = & -\gamma_1 \frac{\ddot{A}+\dot{E}_{\text{min}}}{\big(\dot{A}+E_{\text{min}}\big)^2} - \gamma_2 \frac{\ddot{A} + \dot{E}_{\text{max}}}{\big(\dot{A}+E_{\text{max}}\big)^2}\ .
\end{eqnarray}

Another sensible restriction might be put on the total amount of energy $\mathscr{E}_{\mathrm{max}}$.
This constraint is again implemented via a functional
\begin{equation}
 \label{eq:energy}
 f_2[A] = -\gamma_3 \log\left(\mathscr{E}_{\mathrm{max}}-\mathscr{E}[A]\right) \ ,
\end{equation}
with 
\begin{equation}
 \label{eq:energy1}
 \mathscr{E}[A] = \frac{1}{2}\int\limits_{-T}^{T}\dot{A}^2(t) dt \ .
\end{equation}
Again, this functional will diverge $f_2[A]\to+\infty$ if $\mathscr{E}_{\mathrm{max}}$ is approached.
Consequently, the derivative is given by:
\begin{eqnarray}
 \frac{\delta f_2}{\delta A} & = & -\gamma_3\frac{\ddot{A}}{\mathscr{E}_{max}-\mathscr{E}[A]} \ .
\end{eqnarray}
\newline
\subsection{Parametrized Field Configuration}
\label{sec:parpot}

In certain situations it might be convenient to formulate the optimization problem also on a finite dimensional space.
As an example consider the field configuration (\ref{eq:spulse_sum}) where we fix the functional form but still allow certain parameters to change.

In order to account for such problems we assume that the vector potential is given by a function 
\begin{eqnarray}
 \label{eq:findim_potential}
 A(t)\rightarrow A(t;\mathbf{p})\ ,\quad \mathbf{p}\in\mathbb{R}^n\ ,
\end{eqnarray}
where $\mathbf{p}$ is the set of parameters which are to be optimized. 
For the field configuration (\ref{eq:spulse_sum}), for instance, $\mathbf{p}$ would be the set of field strengths $E_i$, frequencies $\omega_i$ and time lags $t_{0,i}$.
The Lagrangian is then again defined according to
\begin{alignat}{2}
 \label{eq:findim_Lagrangian}
 L(F,&G,H,\mathbf{p},\mu_F,\mu_G,\mu_H)  =  \\
     &J[F,A(\mathbf{p})] + \langle \mu_F,e_F\rangle_{\Omega} + \langle \mu_G,e_G\rangle_{\Omega} + \langle \mu_H,e_H\rangle_{\Omega}\ , \quad \nonumber
\end{alignat}
such that the equations of motion (\ref{eq:eom}) and adjoint equations (\ref{eq:adjoint}) are not changed apart from the replacement $A(t)\rightarrow A(t;\mathbf{p})$. 

Moreover, the reduced cost functional becomes a function of $\mathbf{p}$
\begin{eqnarray}
 \label{eq:findim_red_cost_fct}
 \hat{J}(\mathbf{p}) & = & J[F(A(\mathbf{p})),A(\mathbf{p})]\ .
\end{eqnarray}
Its gradient $\nabla_\mathbf{p} \hat{J}$, which becomes a vector in $\mathbb{R}^n$, can again be obtained from the Lagrangian by evaluating its derivative with respect to $\mathbf{p}$
\begin{widetext}
 \begin{equation}
 \label{eq:findim_gradient} 
 \nabla_\mathbf{p}\hat{J} = \nabla_\mathbf{p}L  =  \nabla_\mathbf{p}f[A(\mathbf{p})] + e\int\limits_{-\infty}^\infty\int\limits_{-T}^T\left\{2\left(\mu_HG-\mu_GH\right)\frac{q-eA}{\omega}\nabla_\mathbf{p} A -\epsilon_{\perp}\left(\mu_GF-\mu_FG-\mu_G\right)\nabla_\mathbf{p}\Big(\frac{\dot{A}}{\omega^2}\Big)\right\}dt dq \ 
 \end{equation}
\end{widetext}
for $F,G,H$ and $\mu_{F},\mu_{G},\mu_{H}$ being the solutions of the equations of motion (\ref{eq:eom}) and the adjoint equations (\ref{eq:adjoint}), respectively. 

Further restrictions on the admissible values of $\mathbf{p}$ are again specified in terms of $f[A(\mathbf{p})]$. 
We use the constraints (\ref{eq:bounds1}) and (\ref{eq:energy}) again to guarantee the boundedness of the field strength and the total energy, respectively. 
For parametrized fields, however, both the vector potential $A(t;\mathbf{p})$ and the electric field $E(t;\mathbf{p})=-\dot{A}(t;\mathbf{p})$ 
depend on the set of parameters $\mathbf{p}$. 
Accordingly, the functionals (\ref{eq:bounds1}) and (\ref{eq:energy}), respectively, also depend on $\mathbf{p}$ such that their derivatives are given by:
\begin{equation}
 \label{eq:findim_bounds}
 \nabla_\mathbf{p}f_1  =  -\int\limits_{-T}^T\left\{\gamma_1\frac{\nabla_\mathbf{p}\dot{A}}{\dot{A}+E_{\text{min}}}+\gamma_2\frac{\nabla_\mathbf{p}\dot{A}}{\dot{A}+E_{\text{max}}}\right\}dt\ ,
\end{equation}
and
\begin{equation}
 \label{eq:findim_energy}
 \nabla_\mathbf{p}f_2  =  \frac{\gamma_3}{\mathscr{E}_{max}-\mathscr{E}[A]}\int\limits_{-T}^T\dot{A} \left\{ \nabla_\mathbf{p}\dot{A} \right\} dt\ .
\end{equation}

\subsection{Optimization}
\label{sec:optimization}

Given the gradient of the reduced cost functional we are in the position to employ standard optimization methods.
The general strategy is to approach a local minimizer of the reduced cost functional iteratively via
\begin{eqnarray}
 p_{k+1} & = & p_k + \alpha_k d_k \ ,
\end{eqnarray}
where $p_k$ are the expansion coefficients of the vector potential, $d_k$ is a search direction and $\alpha_k$ the step size in the $k$-th iteration step.
In the infinite dimensional formulation, $p_k$ corresponds to the values of the vector potential at intermediate times $A(t)$ whereas it represents the vector of parameters $\mathbf{p}$ in the finite dimensional formulation.

The search direction $d_k$ is taken such that the reduced cost functional is improved upon iteration, i.e. driven towards a local minimum.
In general, this condition implies that its inner product with the gradient should be negative.
As a matter of fact, there are various possible choices for the search direction $d_k$ such as taking the negative gradient -- termed the steepest descent method -- or multiplying the gradient with the negative inverse Hessian -- called Newton's method. 
Other possibilities which are usually faster than steepest descent but also more involved include conjugate gradient methods or quasi-Newton methods. 

Finally, the step size $\alpha_k$ is chosen such that the function $\phi_k(\alpha)=\hat{J}(p_k+\alpha d_k)$ fulfills the strong Wolfe conditions \cite{Nocedal1999}
\begin{subequations}
\label{eq:Wolfe}
\begin{eqnarray}
 \phi_k(\alpha_k) & \leq &  \phi_k(0) + c_1 \alpha_k\phi_k'(0)\ ,\\
 |\phi'_k(\alpha_k)| & \leq & c_2 |\phi'_k(0)| \ ,
\end{eqnarray}
\end{subequations}
with $0\leq c_1\leq c_2\leq 1$ at $\alpha_k$. 
Here, the derivative of $\phi_k(\alpha)$ is given by the inner product $\langle \nabla \hat{J}\left(p_k+\alpha d_k\right),d_k\rangle$.
In the infinite dimensional formulation, an $L^2$ inner product on $[-T,T]$ is taken and $\nabla$ represents the functional derivative 
whereas the usual Euclidean inner product on $\mathbb{R}^n$ with $\nabla$ being the gradient is employed for the finite dimensional formulation.

Having specified the search direction $d_k$ --  in our simulations we will employ the steepest descent as well as the Fletcher-Reeves conjugate gradient method -- there remains the task of finding an appropriate step size $\alpha_k$ which fulfills the strong Wolfe conditions (\ref{eq:Wolfe}).
Again, we will employ a simple algorithm here \cite{Nocedal1999}:
Our strategy is to increase the step size $\tilde{\alpha}$ until the interval $[0,\tilde{\alpha}]$ contains admissible steps. 
As soon as this is achieved, we use bisection on this interval.

\section{Results}
\label{sec:results}

In the following we present some results which are based on the optimal control formalism.
For that purpose we initialize the system at a simple field configuration which is then modified according to the optimization procedure.
The equations of motion (\ref{eq:eom}) and the adjoint equations (\ref{eq:adjoint}) are solved with the aid of a Dormand-Prince Runge-Kutta integrator of order 8(5,3).
For simplicity we restrict ourselves to parametrized field configuration of section \ref{sec:parpot}.

\subsection{Necessity of constraints}
\label{sec:res_constraints}

As simple example we start with the Sauter field configuration 
\begin{subequations}
\label{eq:init_conf_sech}
\begin{eqnarray}
 A_\mathrm{in}(t)&=& -\tfrac{E_0}{\omega}\tanh(\omega t) \ , \\
 E_\mathrm{in}(t)&=& E_0 \operatorname{sech}^2(\omega t) \ .
\end{eqnarray}
\end{subequations}
For the moment we focus on the optimization in the one-parameter space $\mathbf{p}=E_0$ whereas we keep $\omega$ fixed.
As initial values we specify $E_0 = 0.1E_\mathrm{cr}$ and $\omega = m/15$.

Upon performing the optimization we observe that the gradient $\nabla_\mathbf{p}\hat{J}$ is indeed directed towards larger values of $E_0$.
However, we cannot find permissible step lengths $\alpha_k$ consistent with the strong Wolfe conditions (\ref{eq:Wolfe}) as the objective functional $\hat{J}(\mathbf{p})$ is not bounded from below.
Accordingly, trying to minimize the objective functional becomes equivalent to taking the limit $E_0\to\infty$.

\begin{figure}[t]
  \centering
  \includegraphics[width=\columnwidth]{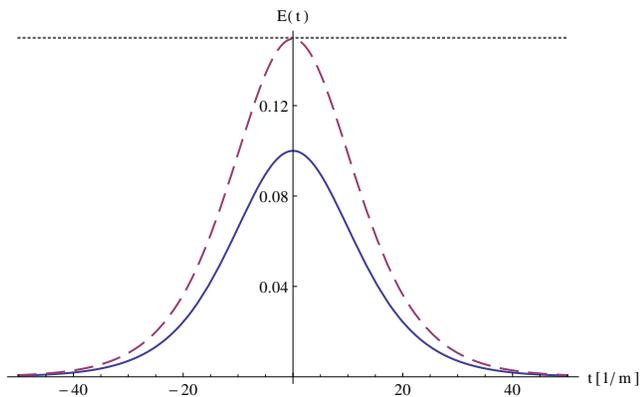}
  \caption{\label{fig:opt_constraints} 
  Electric field configuration:
  The initial configuration (solid) is characterized by $E_0=0.1E_\mathrm{cr}$. 
  After a few iteration steps the field strength $E_0$ converges towards the field strength constraint $E_{0,\mathrm{max}}=0.15E_\mathrm{cr}$.
  The final configuration (dashed) is thus characterized by  $E_0=E_{0,\mathrm{max}}$.}
\end{figure}

In order to set up a convergent optimization procedure we need to include some constraint which prevents the field strength from growing without restraint.
This can be achieved by specifying a functional $f[A(\mathbf{p})]$ which has been introduced in (\ref{eq:cost_functional}) in order to restrict the admissible vector potentials.
Therefore, we start with the same initial configuration (\ref{eq:init_conf_sech}) and employ, for instance, the field strength constraint (\ref{eq:bounds}) with:
\begin{equation}
 \label{eq:fs_const}
  E_{0,\mathrm{max}} = -E_{0,\mathrm{min}}=0.15E_\mathrm{cr} \ .
\end{equation}

As a consequence, the objective functional $\hat{J}(\mathbf{p})$ becomes bounded from below.
The resulting field configuration after a few iteration steps is displayed in FIG.~\ref{fig:opt_constraints}.
As intended, the maximum field strength becomes bounded and the optimization procedure converges.
This shows that it is sometimes absolutely necessary to choose an appropriate constraint in order to obtain sensible optimization results.

Note that we could also employ the energy constraint (\ref{eq:energy}) in order to prevent the electric field strength from increasing without bounds.
This type of constraint is qualitatively similar to a field strength constraint as choosing a maximum energy $\mathscr{E}_\mathrm{max}$ effectively determines a maximal field strength as well.

\subsection{Optimization towards the threshold}
\label{sec:res_threshold}

We consider again the Sauter field configuration (\ref{eq:init_conf_sech}) but this time we fix $E_0=0.1E_\mathrm{cr}$.
Accordingly, we perform an optimization in the one-parameter space $\mathbf{p}=\omega$.
Note that we do not need an additional constraint in this case as the field strength is specified anyway.

In order to perform the optimization, we choose rather different initial values $\omega=m/20$ and $\omega=10m$, respectively.
Upon optimization, we find that the optimal value converges towards $\omega_\mathrm{opt}\sim1.83m$ for both initial values.
As a matter of fact, we may confirm that this value corresponds to the maximum in the particle number near the perturbative threshold by comparison with an analytic result.

\begin{figure}[t]
  \centering
  \includegraphics[width=\columnwidth]{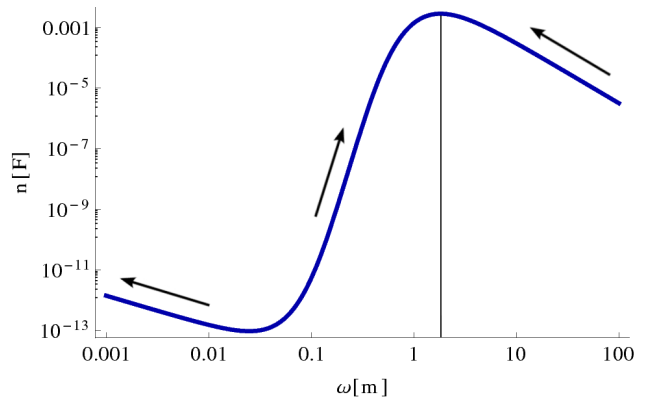}
  \caption{\label{fig:opt_frequency} 
  Particle number $n[F]$ for the Sauter field with $E_0=0.1E_\mathrm{cr}$ as a function of $\omega$. 
  The maximum near the perturbative threshold $\omega_{\mathrm{opt}}\sim1.83m$ is also predicted by our optimal control procedure.
  The arrows indicate the downhill directions of the objective functional $\hat{J}=-\gamma n[F]$.}
\end{figure}

In fact, the asymptotic distribution $F(q,T)$ for the Sauter field is given by \cite{Hebenstreit:2010vz}:
\begin{equation}
 F(q,T)=\frac{2\sinh\big(\tfrac{\pi}{2\omega}\big[2\epsilon+\mu-\nu\big]\big)\sinh\big(\tfrac{\pi}{2\omega}\big[2\epsilon-\mu+\nu\big]\big)}{\sinh(\tfrac{\pi}{\omega}\mu)\sinh(\tfrac{\pi}{\omega}\nu)} \ ,
\end{equation}
with $\epsilon=eE_0/\omega$ and
\begin{equation}
 \mu =\sqrt{\epsilon_\perp^2+(q+\epsilon)^2} \quad , \quad \nu = \sqrt{\epsilon_\perp^2+(q-\epsilon)^2}\ .
\end{equation}
Accordingly, we can explicitly evaluate $n[F]$ which is displayed in FIG.~\ref{fig:opt_frequency}.
Most notably, this curve shows a maximum near the perturbative threshold at $\omega_\mathrm{opt}\sim1.83m$ which coincides with our numerical result.  
For larger values $\omega>\omega_\mathrm{opt}$, the particle number decreases with a power law $\omega^{-2}$.
On the other hand, for smaller values $m/50\lesssim\omega<\omega_\mathrm{opt}$ we observe the characteristic transition between the multiphoton regime to the Schwinger regime which is accompanied by an order-of-magnitude change in the particle numbers.
Moreover, we find an additional maximum at the boundary $\omega\to0$ which is approached with a power law $\omega^{-1}$.
In this limit the Sauter field (\ref{eq:init_conf_sech}) adiabatically changes into a static field $E(t)=E_0$ such that $n[F]$ diverges.

Illustrated by this simple example we want to discuss a numerical challenge which may occur in more complicated situations as well:
Starting with an initial value $\omega \lesssim m/50$ one observes that the local downhill direction of the objective functional $\hat{J}=-\gamma n[F]$ is directed towards $\omega\to0$.
This means that any attempt to start the optimization procedure in this regime should finally drive $\omega\to0$ instead of $\omega\to\omega_\mathrm{opt}$.
Note, however, that the objective functional $\hat{J}$ is rather flat in this regime so that the gradient $\nabla_\mathbf{p}\hat{J}$ is close to numerical precision.
Accordingly, the whole optimization procedure becomes rather untrustworthy.

\subsection{Local maxima vs. global maximum}
\label{sec:res_maxima}

In order to discuss another general issue of optimal control theory, we turn to a more complicated field configuration than the Sauter field (\ref{eq:init_conf_sech}).
To this end we reconsider the field configuration (\ref{eq:spulse_sum}) corresponding to a superposition of arbitrary single pulses.
The need to investigate this kind of field configuration more efficiently was in fact one of our main motivations for the application of the optimal control formalism.

We now re-investigate the comb of $10$ single pulses in both the equal-sign and the alternating-sign configuration:
\begin{subequations}
\begin{eqnarray}
 A_\mathrm{in}(t)&=& -\sum_{i=1}^{10} \tfrac{E_i}{\omega}\tanh(\omega [t-i\cdot t_0]) \ , \\
 E_\mathrm{in}(t)&=& \sum_{i=1}^{10} E_i \operatorname{sech}^2(\omega [t-i\cdot t_0]) \ .
\end{eqnarray}
\end{subequations}
Note again $\operatorname{sign}[E_i]=+1$ for the equal-sign configuration and $\operatorname{sign}[E_i]=\pm1$ in the alternating-sign configuration for $i$ being even and odd, respectively.
Based on our experience we expect the optimization to drive $E_i$ towards higher field strengths and $\omega$ towards the perturbative threshold, respectively.
Accordingly, we fix $|E_i|=0.02E_\mathrm{cr}$ and $\omega=m/6$ such that the the optimization problem is again formulated in a one-dimensional space $\mathbf{p}=t_{0}$ of the inter-pulse time lag.

\begin{table}[b]
  \centering
  \begin{tabular}{c|c|c}
  $\ $ Initial $t_{0,\mathrm{init}}$ $\ $ & $\ $ Optimized $t_{0,\mathrm{opt}}$ $\ $& $\ $Comparison $t_{0,\mathrm{comp}}$ $\ $ \\
  \hline
  $15$& $15.86$& $15.87$ \\
  $17$& $15.75$& $15.87$ \\
  $19$& $18.96$& $18.98$ \\
  $21$& $22.12$& $22.10$ \\
  \end{tabular}
  \caption{\label{tab:opt_time_lag_equ}
  Optimization of the inter-pulse time lag $t_0$ for the equal-sign configuration.
  The values $t_{0,\mathrm{opt}}$ are in good agreement with $t_{0,\mathrm{comp}}$ which are deduced from FIG.~\ref{fig:comb}.} 
\end{table}

The equal-sign configuration is investigated in TAB.~\ref{tab:opt_time_lag_equ} whereas results for the alternating-sign configuration are shown in TAB.~\ref{tab:opt_time_lag_alt}.
We display the initial values $t_{0,\mathrm{init}}$, the corresponding optimized values $t_{0,\mathrm{opt}}$ as well as the optimal values $t_{0,\mathrm{comp}}$ which are deduced from FIG.~\ref{fig:comb}. 
Most notably, we observe that the optimization procedure does not yield a unique optimal value corresponding to a global maximum of $n[F]$ but a whole slew of different optimal values corresponding to different local maxima of $n[F]$.
The reason for this is the fact that optimal control theory deals with the determination of local extrema whereas the identification of the global extremum is beyond its scope.
In order to search for a global maximum of $n[F]$ it is necessary to perform the optimization procedure for various initial values.
In this respect, a proper choice of the initial values can be most crucial: 
A poor choice may put us in the basin of attraction of a 'wrong' minimum of the objective functional $\hat{J}(\mathbf{p})$.

\begin{table}[t]
  \centering
  \begin{tabular}{c|c|c}
  $\ $ Initial $t_{0,\mathrm{init}}$ $\ $ & $\ $ Optimized $t_{0,\mathrm{opt}}$ $\ $& $\ $Comparison $t_{0,\mathrm{comp}}$ $\ $ \\
  \hline
  $16$& $15.43$& $15.46$ \\
  $18$& $17.00$& $17.19$ \\
  $19$& $18.50$& $18.57$ \\
  $20$& $20.50$& $20.30$ \\
  $21$& $21.75$& $21.68$ \\
  $23$& $23.50$& $23.43$ 
  \end{tabular}
  \caption{\label{tab:opt_time_lag_alt}
  Optimization of the inter-pulse time lag $t_0$ for the alternating-sign configuration.
  Note that we resolve both the dominant and the sub-dominant local maxima.}
\end{table}

Moreover, if we compare the optimal values $t_{0,\mathrm{opt}}$ with $t_{0,\mathrm{comp}}$, we observe quite reasonable agreement even though we do not obtain full convergence.
It can only be speculated why this happens, however, it seems that our algorithm tends to considerably slow down once we are close to the local maximum of $n[F]$:
Starting from a given initial value we proceed in very few iteration steps towards this maximum, however, once we are close to it the gradient $\nabla_{\mathbf{p}}\hat{J}$ becomes rather small.
As a consequence, its numerical evaluation becomes error prone such that the optimization procedure is terminated even though full convergence has not yet been achieved.

\subsection{Multi-parameter optimization}

We now turn to two-parameter spaces after discussing only one-parameter optimization problems so far.
In the one-parameter case, the parameter $\mathbf{p}\in\mathbb{R}$ could only be increased or decreased depending on the sign of the gradient $\nabla_\mathbf{p}\hat{J}$.
Upon increasing the dimensionality of the parameter space $\mathbb{R}^n$, however, the gradient becomes an $n$-dimensional vector and, associated with that, several subtleties arise.
Most notably, we will discuss the issue of scaling which is crucial in order to obtain sensible optimization results. 

For simplicity, we restrict ourselves to a two-parameter space in the following.
To be specific, we investigate the double pulse configuration: 
\begin{subequations}
\begin{eqnarray}
 A_\mathrm{in}(t)&=& - \tfrac{E_1}{\omega_1}\tanh(\omega_1 t) - \tfrac{E_2}{\omega_2}\tanh(\omega_2 t) \ , \\
 E_\mathrm{in}(t)&=& E_1 \operatorname{sech}^2(\omega_1 t) + E_2 \operatorname{sech^2}(\omega_2 t) \ .
\end{eqnarray}
\end{subequations}
We keep $\omega_1=m/10$ fixed and demand $E_1+E_2=E_\mathrm{max}$ with $E_\mathrm{max}=0.01E_\mathrm{cr}$.
In this configuration, $E_1$ changes upon variation of $E_2$ such that we are concerned with a two-parameter optimization problem for $\mathbf{p}=(E_2,\omega_2)$.
Note that in order to guarantee $|E(t)| \leq E_\mathrm{max}$ for all $t$, we need to include a field strength constraint (\ref{eq:bounds}) as well.

\begin{table}[t]
  \centering
  \begin{tabular}{l|c|c}
  $\ $ Initial $(E_2,\omega_2)$ $\ $ & $\ $ Optimized $(E_2,\omega_2)$ $\ $& $\ $Optimized $n[F]$ \\
  \hline
  $\quad\ (10^{-4},10)$& $(0.0205,1.830)$& $1.176\cdot10^{-4}$\\
  $\quad\ (10^{-4},1/2)$& $(0.0204,1.827)$& $1.175\cdot10^{-4} $ \\
  $\quad\ (0.005,10)$& $(0.0205,1.819)$& $1.182\cdot10^{-4}$ \\
  $\quad\ (0.005,1/2)$& $(0.0204,1.827)$& $1.175\cdot10^{-4} $ \\
  $\quad\ (0.02,10)$& $(0.0205,1.833)$& $1.181\cdot10^{-4}$ \\
  $\quad\ (0.02,1/2)$& $(0.0204,1.823)$& $1.175\cdot10^{-4}$
  \end{tabular}
  \caption{\label{tab:opt_2d_config} 
  Two-parameter optimization for $\mathbf{p}=(E_2,\omega_2)$ in units of $E_\mathrm{cr}$ and $m$, respectively. 
  We display different initial configurations, the resulting optimized configuration as well as the particle number $n[F]$.}
\end{table}

\begin{figure}[t]
  \centering
  \includegraphics[width=\columnwidth]{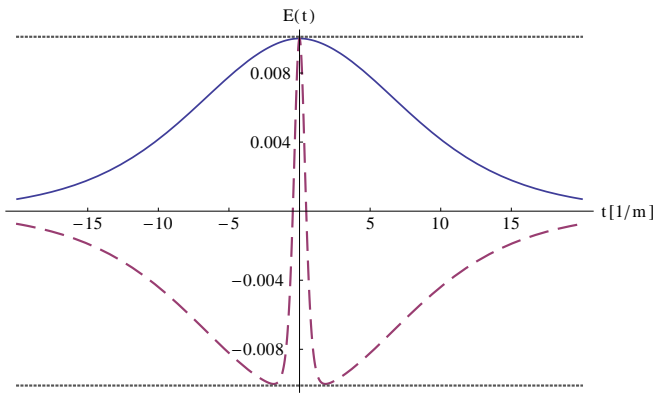}
  \caption{\label{fig:opt_2d_field} 
  Electric field configuration:
  For simplicity we show only one of the initial configurations (solid) which is characterized by $\mathbf{p}_\mathrm{in}=(10^{-4}E_\mathrm{cr},m/2)$.
  The optimized configuration (dashed) converges towards $\mathbf{p}_\mathrm{opt}\sim(0.203E_\mathrm{cr},1.82m)$.}
\end{figure}

In order to perform the optimization, we choose rather different initial configurations as displayed in TAB.~\ref{tab:opt_2d_config}.
In this table one can also see that all these configurations converge quite well towards an optimal configuration $\mathbf{p}_\mathrm{opt}\sim(0.0203E_\mathrm{cr},1.82m)$.
One of the initial configurations as well as the optimized field configuration is displayed in Fig.~\ref{fig:opt_2d_field}.

The behavior of the optimized field configuration is in accordance with our previous investigations of one-parameter problems. 
We have seen that a maximum particle yield is obtained for frequencies which are close to the perturbative threshold.
In accordance with that, the frequency $\omega_2$ is driven towards this threshold $\omega_{2,\mathrm{opt}}\sim1.82m$. 
Moreover, the field strength $E_2$ is maximized subject to the field strength constraint (\ref{eq:bounds}) as well as the condition $E_1+E_2=E_\mathrm{max}$.
This means, on the other hand, that the field strength $E_1<0$ ultimately takes a negative value.

We have already mentioned, that the optimization in a multi-parameter space is associated with several subtleties, most notably the issue of scaling \cite{Nocedal1999}:
Given an \mbox{optimization} problem for many parameters it might happen that the objective functional $\hat{J}(\mathbf{p})$ changes rather differently as a function of the various parameters $\mathbf{p}_i$.
Consequently, the various components of the gradient $\nabla_\mathbf{p}\hat{J}(\mathbf{p})$ may also differ substantially so that the descent direction is dominated by a specific subset of parameters.
If this is the case, the problem is said to be poorly scaled as the rate of convergence towards $\mathbf{p}_\mathrm{opt}$ might decrease significantly.
As a matter of fact, algorithms such as the steepest descent method are very sensitive to poor scaling behavior.
To remedy this shortcoming it is indicated to perform diagonal scaling, i.~e. redefine $\mathbf{p}\to\mathbf{p}'$ such that the objective functional $\hat{J}(\mathbf{p'})$ is better balanced as a function of $\mathbf{p}'$.

\begin{figure}[t]
  \centering
  \includegraphics[width=\columnwidth]{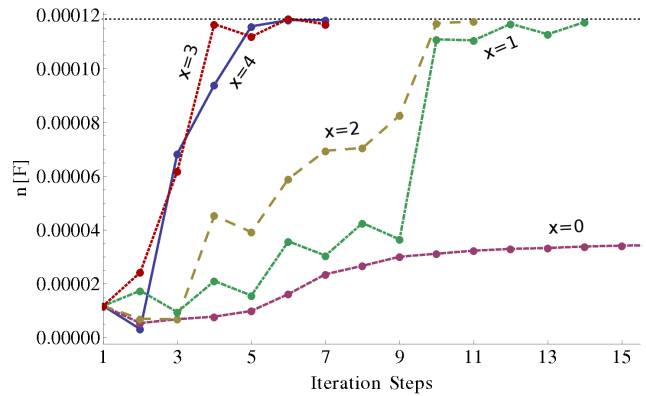}
  \caption{\label{fig:opt_2d_scaling} 
  Rate of convergence of the particle number $n[F]$ for different values of the diagonal scaling exponent $x$ defined in (\ref{eq:scaling_exp}). 
  In each case we start with the initial configuration $\mathbf{p}_\mathrm{in}=(0.02E_\mathrm{cr},10m)$.
  The connecting lines are included to guide the eye.}
\end{figure}

As a matter of fact, our objective functional (\ref{eq:cost_functional}) turns out to be poorly scaled as a function of $\mathbf{p}=(E_2,\omega_2)$.
Accordingly, we need to perform diagonal scaling in order to optimize the convergence properties of our algorithm:
\begin{equation}
 \label{eq:scaling_exp}
 \omega'_2=10^{-x}\omega_2 \ .
\end{equation}
In Fig.~\ref{fig:opt_2d_scaling} we investigate the rate of convergence for a given initial configuration $\mathbf{p}_\mathrm{in}=(0.02E_\mathrm{cr},10m)$ for different values of the diagonal scaling exponent $x$.
We find that the rate of convergence, i.~e. the number of iteration steps which have to be taken until $n[F]$ converges towards its maximum value, strongly depends on the choice of $x$.

As a matter of fact, for $x=0$ the $E_2$-component of the gradient $\nabla_\mathbf{p}\hat{J}(\mathbf{p})$ is orders of magnitude larger than its $\omega_2$-component.
Accordingly, in the course of the optimization the field strength quickly approaches its optimal value $E_{2,\mathrm{opt}}$, however, the optimization of $\omega_2$ does not make much progress.
Due to numerical inaccuracies the $E_2$-component of the gradient  $\nabla_\mathbf{p}\hat{J}(\mathbf{p})$ remains larger than its $\omega_2$-component even if we are close to $E_{2,\mathrm{opt}}$.
Hence, we are finally stuck at a point in parameter space at which $n[F]$ is still way below its optimal value.

On the other hand, we observe that the rate of convergence improves significantly upon increasing the diagonal scaling exponent $x$.
This clearly shows that the issue of scaling is a most crucial one when we want to perform optimization in multi-parameter space.
Most importantly, poor scaling might even prevent the algorithm from converging towards the optimal value.

\section{Conclusion}
\label{sec:conclusion}

Based on optimal control theory, we performed a systematic pulse shaping analysis for electron-positron pair creation in strong external fields.
To this end, we derived the optimal control equations for both finite and infinite dimensional parameter spaces in the framework of quantum kinetic theory.
As this was the first study of this kind, we then focused on the finite dimensional formulation for simplicity.
We convincingly demonstrated that optimal control theory provides a unique means for systematically maximizing the particle yield.
\newpage
The need for pulse shaping has been pointed out previously and recent investigations have again indicated that rather minor parameter changes may increase the particle yield by orders of magnitude \cite{Akkermans:2011yn}.
Due to the fact that a systematic scan of the parameter space becomes rather impracticable soon, it is indicated to perform these pulse shaping analysis by means of an automated tool such as optimal control theory:
Within this framework, the particle yield is driven towards a maximum by an iterative variation of the field configuration.

For the Sauter field configuration, where a comparison with an analytic result is possible \cite{Hebenstreit:2010vz}, we reproduced the trivial maximization directions:
The particle yield is maximized by increasing the field strength at a given frequency or by driving the frequency towards the perturbative threshold at a given field strength.
For more complicated configurations, however, several subtleties arose which will also have to be accounted for in the future:

First, optimal control theory deals with the determination of local maxima of the objective functional whereas the identification of the global maximum is beyond its scope.
Accordingly, a poor choice of initial conditions may put us in the basin of attraction of a 'wrong' maximum.
As it is not possible to comb the whole potential landscape for all local maxima, we still rely on educated guesses or physical intuition for the initial conditions.

Secondly, we have seen that a maximization of the particle yield can always be achieved by increasing the field strength or choosing the frequency near the perturbative threshold.
In order to exclude these trivial maximization directions, we need to apply appropriate constraints.
We demonstrated the effectivity of a field strength constraint convincingly. This serves as an example of optimization towards experimentally realizable configurations.

Finally, we found that a poor scaling behavior is inherent to the objective functional already for a two-parameter space.
As the chosen algorithms are rather sensitive to poor scaling, we performed diagonal scaling.
A probably more elegant way, which will be employed in future investigations, is to apply more advanced methods like non-linear conjugate gradient or approximate Newton methods which are not affected by poor scaling.

Due to the fact that small parameter changes might enhance the particle yield by orders of magnitude, it is surely worthwhile to scan the parameter space for optimal field configurations.
Optimized field configurations could in the long run serve as an input for upcoming high-intensity laser experiments at the Extreme Light Infrastructure (ELI) or the European XFEL.
Eventually, this could facilitate the observation of electron-positron pair creation in the sub-critical field strength regime.
From an experimental point of view, the shaping of femtosecond pulses is an available cutting-edge technique which allows for the experimental generation of complicated field configurations according to user specification \cite{weiner:1929}.
Most prominently, spectral pulse shaping by means of acousto-optic modulators (AOM) is used to modify the phase and the amplitude of femtosecond pulses in a predetermined way \cite{Hillegas:94,Dugan:97}.
This technique has a wide range of applications ranging from nonlinear fiber optics to ultrafast spectroscopy as well as quantum control experiments.
Most notably, it has been demonstrated for atomic ionization, which is rather similar to the Schwinger effect from a theoretical point of view, that the ionization rate is significantly enhanced by an optimized pulse shape \cite{PhysRevLett.92.208301}.

Due to the fact that the results presented in this paper have been based on rather simple parametrized field configurations, a natural next step would be to set up the optimization algorithm for more realistic field configurations.
As a matter of fact, first steps in this direction have already been taken \cite{Kohlfurst:2012}.
In the long run, these studies should provide us with clear predictions on optimized field configurations for which the suppression of the pair production process becomes partially lifted.
Given that such optimal field configurations exist, an experimental verification of non-perturbative electron-positron pair creation seems realistic within the next decades.
 

\subsection*{Acknowledgments}

MM and CK were funded by the Austrian Science Fund, FWF, through the Doctoral Program on Hadrons in Vacuum, Nuclei, and Stars (FWF DK W1203-N16).
FH was supported by the Alexander-von-Humboldt Foundation.
We thank the research core area ``Modeling and Simulation'' for support.

\bibliography{qed_references.bib}

\end{document}